\documentclass[12pt]{article}

\usepackage[dvips]{graphicx}
\usepackage{amssymb}

\newcommand{\be}   {\begin{equation}}
\newcommand{\ee}   {\end  {equation}}
\newcommand{\bea}  {\begin{eqnarray}}
\newcommand{\eea}  {\end  {eqnarray}}
\newcommand{\benn} {\begin{displaymath}}
\newcommand{\eenn} {\end  {displaymath}}
\newcommand{\ba}   {\begin{array}}
\newcommand{\ea}   {\end  {array}}

\newcommand\an[1]{\mbox{$\alpha_s^{#1}$}}
\newcommand\abn[1]{\mbox{$\overline{\alpha}_s^{\,#1}$}}

\newcommand{\vect}[1] {\mbox{\bf #1}}

\newcommand{\gevcc} {\mbox{$\,\mathrm{ GeV/}c^2$}}
\newcommand{\Mz}    {\mbox{${ M}_{\mathrm{ Z}}$}}
\newcommand{\Mzsq}  {\mbox{${ M}^2_{\mathrm{ Z}}$}}
\newcommand{\epem} {\mbox{$e^+e^-$}}
\newcommand{\LEP}  {\mbox{\textsc{LEP}}}
\newcommand{\CERN} {\mbox{\textsc{CERN}}}
\newcommand{\ALE}   {\mbox{\textsc{ALEPH}}}

\newcommand{\DELPHI}{\mbox{\textsc{DELPHI}}}
\newcommand{\OPAL} {\mbox{\textsc{OPAL}}}
\newcommand{\as}  {\mbox{$\alpha_s$}}
\newcommand{\asb} {\mbox{$\overline{\alpha}_s$}}

\newcommand{\nf}  {\mbox{$n_f$}}

\begin{document}


\begin{titlepage}
\vspace{0.5cm}

\renewcommand{\thefootnote}{\fnsymbol{footnote}}

\vspace{1cm}
\flushright{CERN-OPEN-97-015}
\flushright{May 21, 1997}
\flushright{hep-ex/9705016}
\vspace{1.5cm}
\begin{center}
  {\huge\bf Measurements of the\\[0.2cm]
            QCD Colour Factors at 
LEP\footnote[1]{Talk presented at the 5th Topical Seminar on
\textit{The Irresistible Rise of the Standard Model},
San Miniato, Italy, April 1997}}

\vspace{2.5cm}

{\large G. Dissertori}\\[0.3cm]
PPE Division, CERN\\
CH-1211 Geneva 23, Switzerland

\end{center}
\vspace{2cm}

\renewcommand{\thefootnote}{\arabic{footnote}}


\begin{abstract}
\noindent
A summary of the measurements of the QCD colour factors
at \LEP\ is presented. Such measurements provide a test
of the gauge group structure underlying the theory
of strong interactions. A variety of methods have been
applied by the various experiments, and perfect consistency
with the expectation of QCD with SU(3) as gauge group 
is found. 
\end{abstract}

\renewcommand\baselinestretch{1.}
\end{titlepage}


\section{Introduction}
 \label{intro}

The electron-positron collider \LEP\ at \CERN\ has delivered
several million hadronic Z decays to each of the four 
experiments. This large data sample allows for precision
studies of the Standard Model of electroweak and strong 
interactions. In particular it is now possible to perform
precision tests of perturbative 
Quantum Chromodynamics (QCD), the theory
of strong interactions of quarks and gluons. This arises from
the fact that at the high energies at \LEP\ ($E_{cm} = \Mz$)
the strong coupling constant $\as$ is not too large owing to
the property of asymptotic freedom, thus making perturbative
calculations reliable and hadronization effects less
important. Altogether this allows for the observation of hadronic jets,
whose energies and directions are in close  correspondence 
with those of the partons (quarks and gluons), which are not
directly observable.

\begin{figure}[htb]
 \begin{center}
  \includegraphics[bbllx=70,bblly=185,
                   bburx=535,bbury=650,width=10cm]{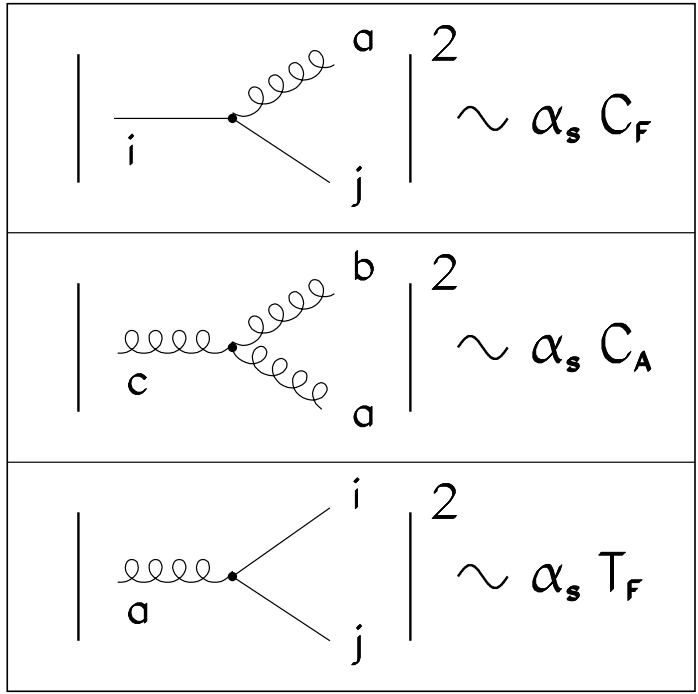}
  \caption{Relations between   vertices and colour factors
           \label{defcol}.}
 \end{center}  
\end{figure}

QCD is a gauge theory with SU(3) as underlying gauge group.
For a  general gauge theory with a simple Lie group, 
the couplings of the fermion
fields to the gauge fields and the self-interactions in the non-abelian
case are determined by the coupling constant and the Casimir operators
of the gauge group. Measuring the eigenvalues of these operators, 
called colour factors,
probes the underlying structure of the theory in a gauge invariant way.
Considering the case where $N_F$ and $N_A$ are  the 
dimensions of the fundamental and adjoint representations of the gauge group
with structure constants $f^{abc}$ and generators $T^a_{ij}$, 
the following relations hold~:
\be
 \label{faftdef}
 \sum_{a=1}^{N_A}\left( T^a T^{\dagger a} \right)_{ij} \mbox{=}\,
  \delta_{ij} C_F \quad ,\quad
 \sum_{i,j=1}^{N_F} T^a_{ij} T^{\dagger b}_{ji} \mbox{=}\,
  \delta^{ab} T_F \quad ,\quad 
 \sum_{a,b=1}^{N_A} f^{abc} f^{*abd} \mbox{=}\, \delta^{cd} C_A ,
\ee
where $a,b,\ldots (i,j,\ldots)$ represent gauge (fermion) field
indices and $C_F$, $C_A$ and $T_F$ are the colour factors.
For  SU($N_C$) one finds
\be
  C_A = N_C \quad , \quad C_F = \frac{N_C^2 - 1}{2 N_C} \quad , 
  \quad T_F=1/2\, .
\ee
For QCD $N_C=3$, hence $C_A = 3$ and $C_F = 4/3$. 
The relations between Feynman vertices and  colour factors 
are illustrated in Fig.~\ref{defcol}. 
In general a cross section for $\epem\rightarrow partons$
has the structure
\be
   \sigma = 
     f(\alpha_s C_F, \frac{C_A}{C_F}, n_f\frac{T_F}{C_F}) \; ,
\ee 
and thus it also depends on the number of active flavours \nf.  
This number could be altered from its LEP 1 
expectation of five by new physics,
such as the existence of a very light gluino \cite{Farrar95}.


\section{Test Variables}
 \label{testvariables}

Basically two different kind of observables have been
used in order to extract information on the colour
factors. They differ in the order at which their 
perturbative prediction starts.


\subsection{First Order Variables}
 \label{fstordervar} 

First order variables  are quantities for which the perturbative
prediction starts at ${\cal O}(\as)$. Examples are event-shape distributions
such as thrust, jet masses, jet broadenings or the differential two-jet rate.
For a general event-shape distribution $y$, which vanishes in the limit of
perfect two-jet topologies, the differential cross section can be written
as \cite{ERT}~: 
\be
 \label{firstorder}
  \frac{1}{\sigma_{tot}} \frac{d \sigma}{d y} =  
  \asb(\mu^2) A(y) \,+\,
  \abn{2}(\mu^2) \,\left( B(y) + b_0 A(y) \ln\frac{\mu^2}{s} \right)
  \,+\, {\cal O}(\abn{3}) \; .
\ee
Here $\sigma_{tot}$ is the total hadronic cross section, and the 
redefinition $\asb(\mu^2) = C_F\as(\mu^2)/(2 \pi)$ of the 
running coupling constant has been adopted. With this 
definition the leading coefficient of 
the QCD beta-function is 
$b_0 = (11/6)\, (C_A/C_F) - (2/3)\, (\nf T_F/C_F)$.

\begin{figure}[tbh]
 \begin{center}
  \includegraphics[bbllx=0,bblly=140,
                   bburx=540,bbury=650,width=10cm]{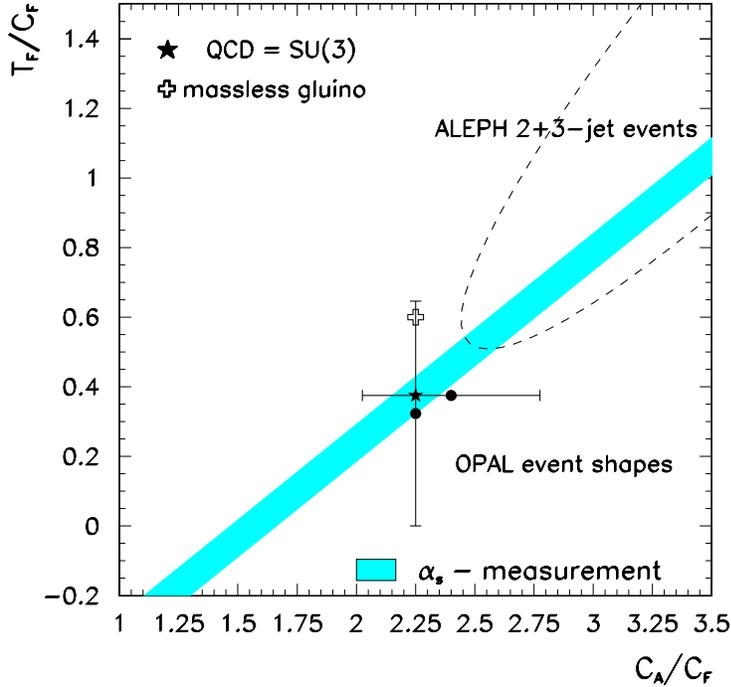}
  \caption{Colour factor measurements based on first order variables
           \label{fstresults}.}
 \end{center}  
\end{figure}

The coefficient functions $A(y)$ and $B(y)$ are obtained by integrating
the fully differential ${\cal O}(\an{2})$
matrix elements \cite{ERT}. Whereas
$A(y)$ is colour factor independent, $B(y)$ can be decomposed into 
\cite{ERT,Magnoli:1990}
\be
  B(y) = B_F(y) + \frac{C_A}{C_F} B_A(y) + \nf\frac{T_F}{C_F} B_T(y) \; .
\ee
From this it becomes clear that information on colour factors
enters only in next-to-leading order via the coefficient function $B(y)$.
However, additional dependence at ${\cal O}(\abn{3})$ and higher orders
enters through the running coupling, mainly via $b_0$, if 
the renormalization scale $\mu^2$ is chosen to be different from
the hard scale $s=E_{cm}^2$.

For several event-shape variables it is possible to resum the leading
and next-to-leading logarithms $\ln y\,$ in all orders of \as\ 
(\cite{Catani:resum-summary} and references therein).
In those cases a function $R(x)$ is added
to the expression in Eq.~(\ref{firstorder}), 
where $x=b_0\, \asb(\mu^2) \ln y$, therefore $b_0$ enters again in 
connection with the leading terms, which introduces a high
correlation between the estimates of $C_A/C_F$ and $\nf T_F/C_F$. 
Summarizing it can be stated that first order variables are suited
for measuring $\as(\Mzsq)$ and a function of $C_A/C_F$ and
$\nf T_F/C_F$, namely $b_0$. This is shown schematically in
Fig.~\ref{fstresults}, where a confidence region in the colour factor
plane is shown for $\as$ determinations from event shapes.

\OPAL\ has used resummation calculations for event shapes to
measure $\as(\Mzsq)$ and one of the colour factors at a time 
\cite{OPcolfac95b}, setting the
others to the QCD expectations. The results are shown in 
Fig.~\ref{fstresults}, together with an \ALE\ measurement
\cite{ALEMegapaper} where a second-order calculation for the 
differential three-jet cross section has been used. 
As they extracted additional
information from the two-jet rate, it was possible to measure
simultaneously $\as$ and both colour factor ratios. However,
the systematic errors are large, mainly owing to the large
renormalization scale uncertainty when using only second-order
calculations.

\begin{figure}[htb]
 \begin{center}
  \includegraphics[bbllx=0,bblly=140,
                   bburx=540,bbury=650,width=10cm]{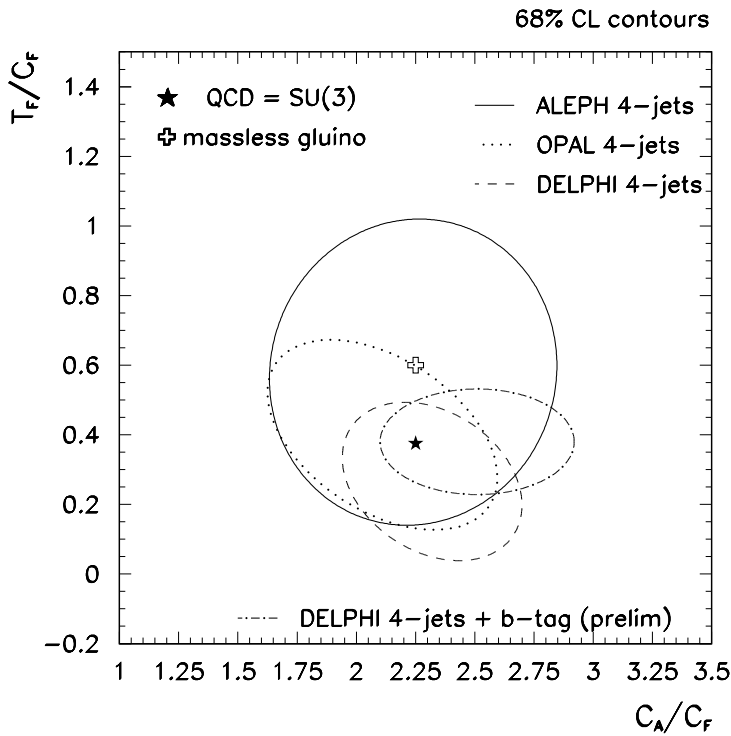}
  \caption{Colour factor measurements based on second order variables
           \label{4jetsummary}.}
 \end{center}  
\end{figure}


\subsection{Second Order Variables}
 \label{sndordervar} 

Information on the colour factors
can also be obtained from second order variables, for which 
the perturbative expansion starts only at ${\cal O}(\an{2})$, like
thrust-minor, light jet mass or angular variables in four-jet events.
Here the cross section is given by   
\be
 \label{secondorder}
  d \sigma \propto 
  \abn{2} \left( D_F +  \frac{C_A}{C_F} D_A + 
  \nf\frac{T_F}{C_F} D_T \right) ,
\ee
\noindent 
with coefficient functions $D_*$ ($*=F,A,T$) again found by integration of
the ${\cal O}(\an{2})$ matrix elements.
Such variables have the advantage that
colour factor information already enters  in leading order, but normally
they are not used to measure \as, as only Born level calculations
were available until recently. Hence in the case of four-jet angular variables
only the shapes of distributions were fitted.
First results of a next-to-leading order
calculation \cite{Dixon:privat} indicate that these shapes remain almost
unchanged under inclusion of higher orders, although, the four-jet rate
is changed.
Like in the case of  first order variables,
a light gluino would alter the cross section.

Four-jet angular distributions \cite{4jetvariables}
 have proven to be very sensitive to the
colour factors. First, four-jet events are found according
to a given clustering prescription and resolution criterion.
If the primary quarks are tagged via energy ordering
($E_1 > E_2 > E_3 > E_4$), the following variables can then be computed
($\vect{p}_i$ being the jet momenta)~:
\begin{itemize}
  \item the Bengtsson-Zerwas angle~:\\
        $\chi_{BZ} = \measuredangle [ (\vect{p}_1 \times \vect{p}_2),
                              (\vect{p}_3 \times \vect{p}_4) ]$
  \item the K\"orner-Schierholz-Willrodt angle~:\\
        $\Phi_{KSW} = 1/2\, \{ \measuredangle [ (\vect{p}_1 \times \vect{p}_4),
                                        (\vect{p}_2 \times \vect{p}_3) ] +
                               \measuredangle [ (\vect{p}_1 \times \vect{p}_3),
                                        (\vect{p}_2 \times \vect{p}_4) ] \}$
  \item the (modified) Nachtmann-Reiter angle~:\\
        $\theta^*_{NR} = \measuredangle [ (\vect{p}_1 - \vect{p}_2),
                                  (\vect{p}_3 - \vect{p}_4) ]$
  \item the angle between the two lowest
         energetic jets~:
        $\alpha_{34} = \measuredangle [ \vect{p}_3, \vect{p}_4 ]$
\end{itemize}

\DELPHI\ \cite{DEcolfac93}
has performed a least-squares fit of Eq.~(\ref{secondorder})
to the two-dimensional distribution in the variables $\theta^*_{NR}$
and $\alpha_{34}$ in order to find estimates of the colour factor ratios.
A similar technique was applied by \OPAL, there, however, a
three-dimensional distribution was measured by using also
the angle $\chi_{BZ}$. 
Recently it has been shown by \DELPHI\ \cite{DEcolfacbtag:96} that the
sensitivity of these angular distributions to the colour factors  
can be further improved by tagging two of the four jets as originating
from b- or c-quarks, using lepton $p_\perp$ and lifetime information.
This method gives an efficiency of about $12 \%$
to tag both primary jets correctly
 and a purity of $70 \%$, whereas with energy ordering 
only in $42\%$ of all events do the two most energetic jets originate from
the primary quarks. 

\begin{figure}[bth]
 \begin{center}
  \includegraphics[bbllx=90,bblly=225,
                   bburx=510,bbury=720,width=10cm]{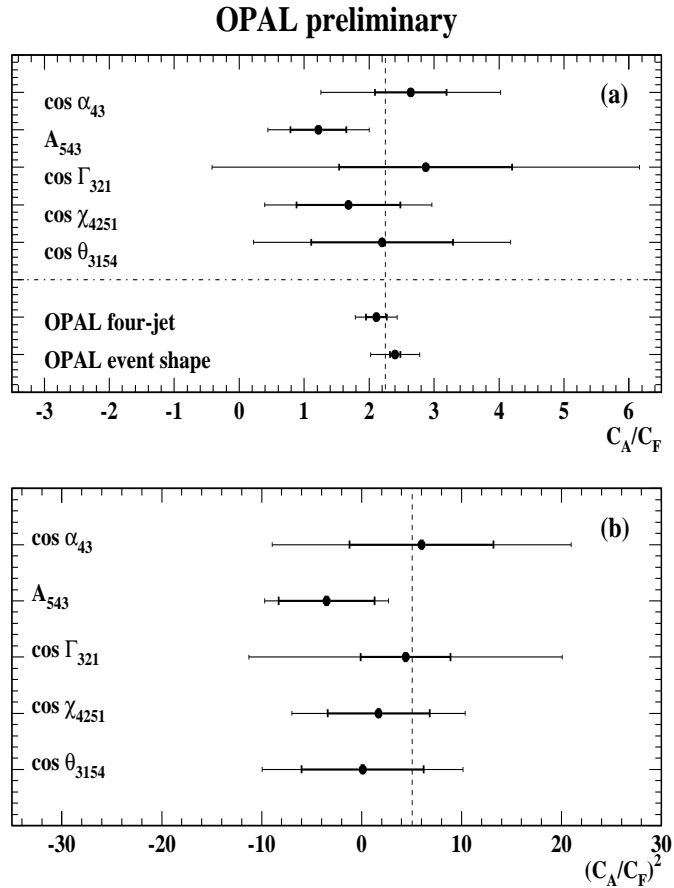}
  \caption{Measurements of $C_A$ and $C_A^2$, which is related to the 
           quartic  gluon vertex
           \label{opal_5jetresult}.}
 \end{center}  
\end{figure}

\ALE\ \cite{ALEcf92} has tried a different approach which is based
on a maximum  likelihood fit of selected four-jet events to the 
theoretical prediction for the five-fold differential four-jet cross
section.

The results of the measurements mentioned above are summarized in
Fig.~\ref{4jetsummary}. Most of them were
statistically limited. The systematic errors
are typically dominated by hadronization uncertainties and systematic
effects related to two- and three-jet background.

\OPAL\ has made a first attempt to study angular correlations in
five-jet events \cite{OPfivejet:1995} in order to determine 
the fraction of the cross section proportional to $C_A^2$, which is the
colour class that contains the quartic gluon vertex diagrams and 
enters only at ${\cal O}(\an{3})$. The results are summarized in 
Fig.~\ref{opal_5jetresult}. Consistency with the expectation 
from QCD is found, however, the errors are large.


\section{Combination of First and Second Order Variables}
 \label{combination}

\begin{figure}[bht]
 \begin{center}
  \includegraphics[bbllx=10,bblly=90,
                   bburx=380,bbury=450,width=11cm]{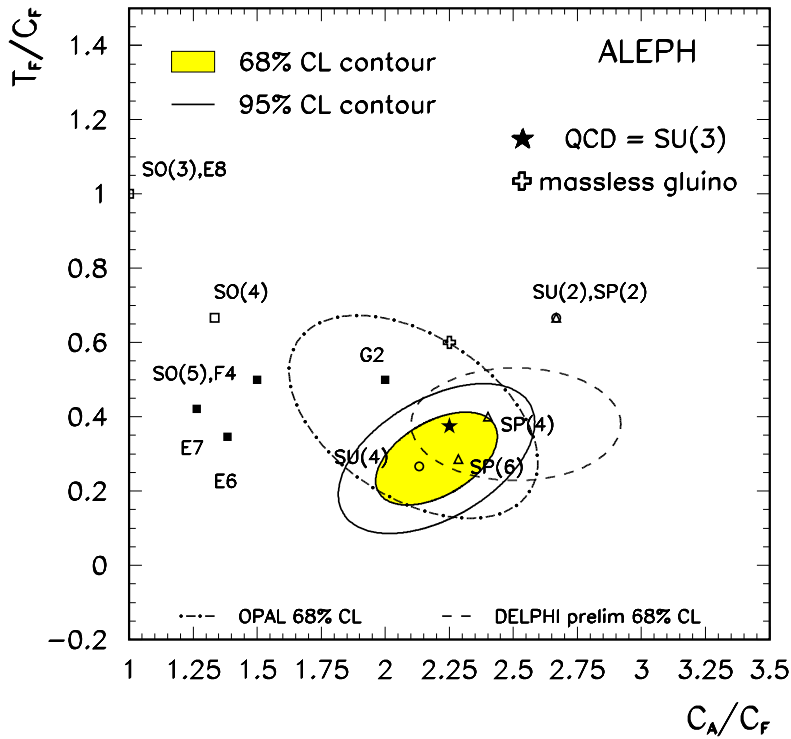}
  \caption{Results of the colour factor measurement by \ALE, compared
           to measurements of \OPAL\ and \DELPHI. Also indicated are
           the expectations from SU(3) and other gauge groups 
           \label{faft_full}.}
 \end{center}  
\end{figure}

Recently \ALE\ has presented a simultaneous measurement of the
strong coupling constant and the colour factors \cite{ALE:colfact-gluino}.
The idea of this analysis is to combine information on these
parameters obtained both from first and second order variables.
As first order variable the distribution in $-\ln y_3$ has been
employed, where $y_3$ is the minimum distance scale $y_{ij}$,
computed according to the Durham prescription 
\cite{Catani:resum-jet-rates,durham3},
after clustering an event to three jets.
This variable
is also called differential two-jet rate, and the  resummation
of leading and next-to-leading logarithms is available for it
\cite{Catani:resum-jet-rates,DissSchmell95}. 
As second order variables,
all four angular distributions defined in Section \ref{sndordervar}
have been used, with jets ordered in energy. In total,
2.7 million hadronic events have been analyzed, giving a large
sample of about 170000 four-jet events. A least-squares
fit has been performed simultaneously to all five one-dimensional
distributions by taking into account the correlations. The fit
parameters  were $\asb(\Mzsq)$, $C_A/C_F$ and $\nf T_F/C_F$. From 
an analysis of mass effects on the four-jet variables it is concluded
that the effective number of flavours is $n_f^{eff}=5.67$, which is then
used to extract the final results for the colour factor ratios,
\benn
 \frac{C_A}{C_F}  =  2.20 \pm 0.09\, (stat) \pm  0.13\, (syst) \quad ,\quad
 \frac{T_F}{C_F}  =  0.29 \pm 0.05\, (stat) \pm  0.06\, (syst) . 
\eenn
These represent the most precise measurements to date. 
The systematic uncertainties include  contributions from renormalization
scale and matching ambiguity (matching of fixed order 
and resummation calculation), from hadronization effects and 
biases introduced via the detector simulation, and finally from
the estimation of mass effects.
The result is illustrated in Fig.~\ref{faft_full} and compared to
the best measurements by \OPAL\ and \DELPHI.

From Fig.~\ref{faft_full} it becomes already clear that the
light gluino hypothesis is not favoured by the colour factor
measurements. A further test of this hypothesis is obtained by
fixing the colour factors to their SU(3) values, thus assuming 
QCD to be the correct gauge theory of strong interactions,
and instead extracting $\as(\Mzsq)$ and $\nf$ from a fit to
the differential two-jet rate and angular distributions. The results
are $
  \as(\Mzsq) =  0.1162 \pm 0.0012\, (stat) \pm 0.0040\, (syst)
$ and $
  \nf      =    4.24 \pm 0.29\, (stat) \pm 1.15\, (syst) 
$.
The measurement of $\as(\Mzsq)$
is in agreement with the world average of $0.118\pm 0.003$
\cite{Schmelling:Warsaw}, and the result for \nf\ is consistent
with the expectation of five. At leading order a massless gluino would  
lead to an excess above five of three units. However, mass effects
can lower this excess. From their measurement, \ALE\ computes
an upper limit on the excess of $\Delta n_f < 1.9$ at $95 \%$ CL, from
which they deduce a lower limit on the gluino mass of
$m_{\tilde g} > 6.3 \gevcc$. Similar conclusions are drawn by
the analyses of the running of \as\ of Refs.~\cite{Schmelling-Denis:1994} 
and \cite{Fodor:96}. There the measurements of $R_{\tau}$ and
$R_{Z}$ are used in order to measure $\as$ at different scales and
thus to test the running coupling, which is a function of the
colour factor ratios and the number of active flavours, as shown
in Section \ref{intro}.


\section{Conclusions}
 \label{conclusio}

Measurements of the QCD colour factors have been performed by
the \LEP\ experiments, based on a variety of approaches.
Consistency with the expectations from QCD with SU(3) as 
gauge group has been found in all cases. The existence of a
very light gluino is ruled out, which however has yet to be 
confirmed by direct searches.


\section{Acknowledgements}

I would like to thank J.~Fuster, A.~Seitz,
S.~Bentvelsen and S.~Kluth for providing me with relevant
information for the preparation of this talk.


\providecommand{\bysame}{\leavevmode\hbox to3em{\hrulefill}\thinspace}

\end{document}